\documentclass[aps,pra,twocolumn,groupaddresses,superscriptaddress,showpacs]{revtex4}
\usepackage{times}
\usepackage{latexsym}
\usepackage{graphicx}
\usepackage{amsmath,amssymb}
\usepackage{verbatim,bbm}

\newtheorem{theorem}{Theorem}
\newtheorem{lemma}[theorem]{Lemma}

\input{tcilatex}

\begin{document}

\title{Ultimate precision bound of quantum and sub-wavelength imaging}
\author{Cosmo Lupo}
\affiliation{York Centre for Quantum Technologies (YCQT), University of York, York YO10
5GH, UK}
\author{Stefano Pirandola}
\affiliation{York Centre for Quantum Technologies (YCQT), University of York, York YO10
5GH, UK}
\affiliation{Computer Science, University of York, York YO10 5GH, UK}

\begin{abstract}
We determine the ultimate potential of quantum imaging for boosting the
resolution of a far-field, diffraction-limited, linear imaging device within
the paraxial approximation. First we show that the problem of estimating the
separation between two point-like sources is equivalent to the estimation of
the loss parameters of two lossy bosonic channels, i.e., the
transmissivities of two beam splitters. Using this representation, we
establish the ultimate precision bound for resolving two point-like sources
in an arbitrary quantum state, with a simple formula for the specific case
of two thermal sources. We find that the precision bound scales with the
number of collected photons according to the standard quantum limit. Then we
determine the sources whose separation can be estimated optimally, finding
that quantum-correlated sources (entangled or discordant) can be
super-resolved at the sub-Rayleigh scale. Our results set the upper bounds
on any present or future imaging technology, from astronomical observation
to microscopy, which is based on quantum detection as well as source
engineering.
\end{abstract}

\maketitle

%---

\textit{Introduction.} Quantum imaging aims at harnessing quantum features
of light to obtain optical images of high resolution beyond the boundary of
classical optics. Its range of potential applications is very broad, from
telescopy to microscopy and medical diagnosis, and has motivated a
substantial research activity~\cite%
{Helstrom,Fabre,sub-R,Tsang09,Boyd,Exp1,Exp2,Exp3,Exp4,Tsang}. Typically,
quantum imaging is scrutinized to outperform classical imaging in two ways.
First, to resolve details below the Rayleigh length (sub-Rayleigh imaging).
Second, to improve the way the precision scales with the number of photons,
by exploiting non-classical states of light. It is well known that a
collective state of $N$ quantum particles has an effective wavelength that
is $N$ times smaller than individual particles \cite%
{1992,1994,1996,N00N,2004,Paris,Rafal,review}. If $N$ independent photons
are measured one expects that the blurring of the image scales as $1/\sqrt{N}
$ (known as \textit{standard quantum limit} or \textit{shot-noise limit}),
while for $N$ entangled photons one can sometimes achieve a $1/N$ scaling
(known as the \textit{Heisenberg limit}).

In this Letter we compute the optimal resolution limit of quantum imaging
for estimating the linear or angular separation between two point-like
monochromatic sources, by using a linear diffraction-limited imaging device
in the far-field regime and within paraxial approximation \cite{Goodman}. In
this way, we determine the ultimate capabilities of quantum light for
boosting the resolution of optical imaging, setting the upper bound on any
present or future imaging technology. We show that the ultimate precision
bound scales with the number of photons according to the standard quantum
limit, for an arbitrary state of the sources. We then study the precision
achievable for sources which are in thermal, discordant or entangled states.
We determine the optimal entangled states that saturate the bound and we
show that sources of quantum-correlated light yield optimal imaging of
sub-Rayleigh features, allowing for higher resolution below the Rayleigh
length. Our findings generalize the seminal Ref.~\cite{Tsang} which has led to
several experimental advances in quantum imaging~\cite{e1,e2,e3,e4}.

\begin{figure}[t]
\centering
\includegraphics[width=0.4\textwidth]{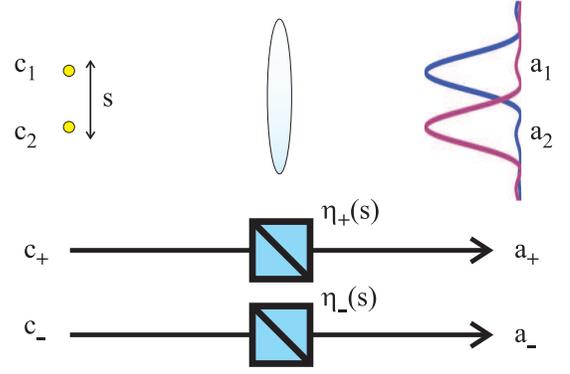}
\caption{ A diffraction-limited linear-optical system creating an image of
two point-like sources (top of the figure) is formally equivalent to a pair
of independent beamsplitters (bottom of the figure), whose transmissivities
are functions of the separation between the sources, with $c_\pm = (c_1 \pm
c_2)/\protect\sqrt{2}$.}
\label{fig:model}
\end{figure}

To achieve our results, we estimate the ultimate precision bound in terms of
the quantum Fisher information. The ultimate error of any unbiased estimator
of the separation $s$ between two sources is given by the quantum Cram\'{e}r-Rao
bound~\cite{1994,1996} 
\begin{equation}
\Delta s\geq \frac{1}{\sqrt{\mathrm{QFI}_{s}}}\,,
\end{equation}%
where $\mathrm{QFI}_{s}$ is the quantum Fisher information. The latter is a
function of $s$, of the features of the optical imaging system, and of the
state of the light emitted by the sources. Here we show that a linear
diffraction-limited imaging system in the paraxial approximation is
equivalent to a pair of beam splitters, whose transmissivities are functions
of the separation (see Fig.\ \ref{fig:model}). Thus, we reduce the estimate
of the separation to the estimate of the transmissivity of a beam splitter~%
\cite{BS1,BS2,Pinel,Venzl,Gaiba,Monras,Gae}. In this way, not only we are
able to compute the quantum Fisher information for any pair of sources but
we also determine the optimal sources that saturate the ultimate precision bound.

%---

\textit{The quantum model.} Consider the canonical annihilation and
creation operators, $c_{1},c_{1}^{\dag }$ and $c_{2},c_{2}^{\dag }$,
describing two monochromatic point-like sources. The sources are separated
by distance $s$ and lay on the object plane orthogonal to the optical axis
at position $-s/2$ and $s/2$. The imaging system maps the source operators
into the image operators $a_{1},a_{1}^{\dag }$ and $a_{2},a_{2}^{\dag }$,
describing the optical field on the image screen. Without loss of
generality, we assume that the optical system has unit magnification factor.
This implies that the point spread function has the form $T(x,y)=\sqrt{\eta }%
\,\psi (x-y)$, where $x$ and $y$ are respectively the coordinates on the
image and object plane, $\psi $ is a function on the image plane with unit $%
L_{2}$-norm, and $\eta $ is an attenuation factor. In particular, the image
operators read 
\begin{align}
a_{1}^{\dag }& =\int \,dx\,\psi (x+s/2)\,a_{x}^{\dag }\,,  \label{defa1} \\
a_{2}^{\dag }& =\int \,dx\,\psi (x-s/2)\,a_{x}^{\dag }\,,  \label{defa2}
\end{align}%
where $a_{x}$, $a_{x}^{\dag }$ are the canonical creation and annihilation
operators for the field at location $x$ on the image screen.

The image modes are distorted and attenuated versions of the source modes.
In fact, the optical imaging system transforms the source operators as \cite{NOTAnew}
\begin{align}
c_{1}& \rightarrow \sqrt{\eta }\,a_{1}+\sqrt{1-\eta }\,v_{1}\,, \\
c_{2}& \rightarrow \sqrt{\eta }\,a_{2}+\sqrt{1-\eta }\,v_{2}\,,
\end{align}%
where $v_{1}$, $v_{2}$ are auxiliary environmental modes that we will assume
to be in the vacuum state (this is a physically reasonable assumption at
optical frequencies). Because of the non-zero overlap between the two point
spread functions $\psi (x+s/2)$ and $\psi (x-s/2)$, the\ image operators $%
a_{1}$ and $a_{2}$ are not orthogonal, i.e., they do not satisfy the
canonical commutation relations. % $[a_{1},a_{2}^{\dag }]=0$.
In order to make them orthogonal we take the sum and difference of the above
relations, obtaining 
\begin{align}
c_{+}& :=\frac{c_{1}+c_{2}}{\sqrt{2}}\rightarrow \sqrt{\eta _{+}}\,a_{+}+%
\sqrt{1-\eta _{+}}\,v_{+}\,,  \label{c+} \\
c_{-}& :=\frac{c_{1}-c_{2}}{\sqrt{2}}\rightarrow \sqrt{\eta _{-}}\,a_{-}+%
\sqrt{1-\eta _{-}}\,v_{-}\,,  \label{c-}
\end{align}%
where $\eta _{\pm }:=(1\pm \delta )\eta $ are transmissivities depending on
the image overlap 
\begin{equation}
\delta =\mathrm{Re}\int dx\,\psi ^{\ast }(x+s/2)\psi (x-s/2)
\end{equation}%
between the non-orthogonal modes $a_{1}$ and $a_{2}$, and 
\begin{equation}
a_{\pm }:=\frac{a_{1}\pm a_{2}}{\sqrt{2(1\pm \delta )}}  \label{sym-image}
\end{equation}%
are orthogonal symmetric and antisymmetric canonical operators on the image
plane.

The non-local source modes $c_{\pm }$ are hence independently mapped and
attenuated into the image modes $a_{\pm }$, by means of effective
attenuation factors $\eta _{\pm }=(1\pm \delta )\eta $, as also shown in 
Fig.~\ref{fig:model}. 
Inverting Eqs.\ (\ref{c+})-(\ref{c-}) we write
\begin{equation}\label{BSmain}
a_{\pm }=\sqrt{\eta _{\pm }}\,c_{\pm }-\sqrt{1-\eta _{\pm }}\,v_{\pm } \, .
\end{equation}%
Note that the overlap $\delta $
between the two point spread functions is a crucial parameter in our model:
it quantifies the diffraction introduced by the imaging optical system, as
well as the amount of constructive (destructive) interference in the
symmetric (antisymmetric) image modes. Also note that this model is
well-defined only for $\eta \leq 1/2$: we remark that this is in accordance
with the fact that in the paraxial approximation a point source is always
(by definition) imaged in the far-field regime, in which light is attenuated
by a factor $\eta \ll 1$ (see, e.g., Refs.~\cite{Shapiro,mio})~\cite{NOTA1}. 

Our equivalent representation of the imaging process leads to a simple
description for the dynamical evolution of the image operators $a_{\pm }$ in
terms of the separation $s$ between the source. In fact, we may prove the
following.

\begin{lemma}
\label{Lemma1} Consider a diffraction-limited linear-optical system creating
an image of two point-like sources. The symmetric and antisymmetric image
operators $a_{\pm }$ satisfy the following dynamical equations in terms of
the separation parameter%
\begin{equation}
\frac{da_{\pm }}{ds}=i\omega _{\pm }[H_{\pm }^{\mathrm{eff}},a_{\pm }]\,,
\label{LemmaEQ}
\end{equation}%
where $H_{\pm }^{\mathrm{eff}}$ are suitable beam splitter-like Hamiltonians
and $\omega _{\pm }$ are suitable angular frequencies.
\end{lemma}

\textbf{Proof.}~~First of all, from Eq.~(\ref{BSmain}) we may equivalently
write $a_{\pm }=e^{i\theta _{\pm }H_{\pm }}c_{\pm }\,e^{-i\theta _{\pm
}H_{\pm }}$, where 
\begin{equation}
H_{\pm }=i(c_{\pm }^{\dag }v_{\pm }-v_{\pm }^{\dag }c_{\pm })  \label{H1}
\end{equation}%
are two independent beam splitter-like Hamiltonians with rotation angles $%
\theta _{\pm }=\arccos \sqrt{\eta _{\pm }}$~\cite{FOP}. In the Heisenberg
picture, the dynamics with respect to $s$ is therefore expressed by%
\begin{equation}
\frac{da_{\pm }}{ds}=i\frac{d\theta _{\pm }}{ds}[H_{\pm },a_{\pm }]+\frac{%
\partial a_{\pm }}{\partial s}\,.  \label{dynamics-a}
\end{equation}

The term with the commutator already describes an effective beam
splitter-like transformation. It remains to analyze the term with the
partial derivative. After some derivation, we found that the operators 
$\frac{\partial a_{\pm }}{\partial s}$ are proportional to a pair of
corresponding canonical operators $b_{\pm }$ 
(see Appendix \ref{App_A} for details and exact definitions). 
We obtain 
\begin{equation}
\frac{\partial a_{\pm }}{\partial s}=-\frac{\epsilon _{\pm }}{2\sqrt{1\pm
\delta }}\,b_{\pm }\,,  \label{partial-a}
\end{equation}%
with 
\begin{align}
\epsilon _{\pm }^{2}& =\Delta k^{2}\mp \beta -\frac{\gamma ^{2}}{1\pm \delta 
}\,,  \label{epsilon} \\
\Delta k^{2}& :=\int dx\left\vert \frac{d\psi (x)}{dx}\right\vert
^{2},~~\gamma :=\frac{d\delta }{ds},  \label{dk2} \\
\beta & :=\int dx\frac{d\psi (x+s/2)}{dx}\,\frac{d\psi (x-s/2)}{dx}\,.
\label{beta2}
\end{align}%

From Eqs.\ (\ref{H1}), (\ref{dynamics-a}) and (\ref{partial-a}) we obtain 
\begin{equation}
\frac{da_{\pm }}{ds}=-\frac{d\theta _{\pm }}{ds}\,v_{\pm }-\frac{\epsilon
_{\pm }}{2\sqrt{1 \pm \delta }}\,b_{\pm }\,,
\end{equation}%
so that the dynamical equations can be locally written as in Eq.~(\ref%
{LemmaEQ}), where the angular frequencies are given by 
\begin{equation}
\omega _{\pm }=\sqrt{\left( \frac{d\theta _{\pm }}{ds}\right) ^{2}+\frac{%
\epsilon _{\pm }^{2}}{4(1\pm \delta )}}\,,
\end{equation}%
and $H_{\pm }^{\mathrm{eff}}=i\left( c_{\pm }^{\dag }d_{\pm }-d_{\pm }^{\dag
}c_{\pm }\right) $ are beam splitter-like Hamiltonians mixing the source
modes $c_{\pm }$ with the auxiliary modes 
\begin{equation}
d_{\pm }=\frac{1}{\omega _{\pm }}\left[ \frac{d\theta _{\pm }}{ds}\,v_{\pm }+%
\frac{\epsilon _{\pm }}{2\sqrt{ 1 \pm \delta }}\,b_{\pm }\right]%
.~\blacksquare
\end{equation}

Note that the parameters $\epsilon _{\pm }^{2}$ in Eq.\ (\ref{epsilon})
contain three terms: (i)$%
~\gamma ^{2}$ accounts for the variations of the overlap $\delta $ due to
changes of the separation $s$; (ii)~$\Delta k^{2}$ equals the variance of 
the momentum operator $- i \frac{d}{d x}$ and hence describes translations on
the image screen; and (iii) $\beta $ accounts for interference between the
derivatives of the point spread functions.

%---

%---

\textit{Upper bound on the quantum Fisher information.} With Lemma \ref%
{Lemma1} we have shown that estimating the separation between the sources is
equivalent to estimating the angle of rotation of a beam splitter-like
transformation. We now obtain the fundamental limits of quantum and
sub-Rayleigh imaging by exploiting the fact that the quantum Fisher
information for the angle of a beam splitter rotation, when the other input
port of the beam splitter is in the vacuum state, is no larger than $4\bar{n}
$, where $\bar{n}$ is the mean photon number~\cite[page~4]{BS1}.
In our setting, the fact that the other beam splitter port is in the 
vacuum state corresponds to the assumption that the only light entering the
optical system is that coming from the sources to be imaged, i.e., we
are neglecting any source of background radiation,
which is a natural assumption at optical frequencies.

\begin{theorem}
\label{Theorem1} Consider two point-like sources with unknown separation $s$%
, and emitting a total of $2N$ mean photons, which are observed by an
optical system with point spread function $T(x,y)=\sqrt{\eta }\,\psi (x-y)$
and attenuation $\eta $. Then, the quantum Fisher information cannot exceed
the upper bound%
\begin{equation}
\mathrm{QFI}_{s}\leq \frac{2\eta N}{\mathsf{x_{R}}^{2}}\max \left\{
f_{+}\,,f_{-}\right\} \,,  \label{UB-new}
\end{equation}%
where $\mathsf{x_{R}}$ is the Rayleigh length and the $f$-functions are
given by $f_{\pm }:=\mathsf{x_{R}}^{2}\{\epsilon _{\pm }^{2}+\gamma
^{2}(1\pm \delta )^{-1}[1-(1\pm \delta )\eta ]^{-1}\}$.%
\end{theorem}

\textbf{Proof.}~~To obtain the upper bound, assume that we can measure not
only the image modes $a_{\pm }$, but also the vacuum modes $v_{\pm }$, $%
b_{\pm }$. Let us denote as $|\psi \rangle _{c_{+}c_{-}}$ the state of the
light emitted by the sources $c_{\pm }$. The state of the light at the image
screen, together with the state of the auxiliary modes $v_{\pm }$ and $%
b_{\pm }$ is 
\begin{equation}
|\psi ^{\prime }\rangle =\left( e^{-i\theta _{+}H_{+}}e^{-i\theta
_{-}H_{-}}|\psi \rangle _{c_{+}c_{-}}|0\rangle _{v_{+}v_{-}}\right)
|0\rangle _{b_{+}b_{-}}\,.  \label{psi1}
\end{equation}

According to Lemma~\ref{Lemma1}, the dynamics with respect to $s$ is
described by the effective beam splitter Hamiltonian $H^{\mathrm{eff}%
}=\omega _{+}H_{+}^{\mathrm{eff}}+\omega _{-}H_{-}^{\mathrm{eff}}$. The
upper bound on the quantum Fisher information is therefore obtained by the
formula~\cite{1994,1996}%
\begin{equation}
\mathrm{QFI}_{s}\leq 4\langle \psi ^{\prime }|\Delta ^{2}H^{\mathrm{eff}%
}|\psi ^{\prime }\rangle \,.
\end{equation}%
The calculation of the right hand side of this inequality is reported in
Appendix \ref{app:variance}, where we use the upper bound of Ref.~\cite{BS1}. In
particular, if source $c_{\pm }$ emits $N_{\pm }$ mean photons, i.e., $%
\langle \psi |c_{\pm }^{\dag }c_{\pm }|\psi \rangle =N_{\pm }$, then we
obtain 
\begin{equation}
\mathrm{QFI}_{s}\leq \frac{\eta }{\mathsf{x_{R}}^{2}}\left(
N_{+}f_{+}+N_{-}f_{-}\right) \,.
\end{equation}%
Now, if we fix the total number of photons $2N=N_{+}+N_{-}$, then the
maximum is obtained by either $(N_{+},N_{-})=(2N,0)$ or $(N_{+},N_{-})=(0,2N)
$, yielding the bound of Eq.~(\ref{UB-new}).$~\blacksquare $

\begin{figure}[t]
\centering
\includegraphics[width=0.45\textwidth]{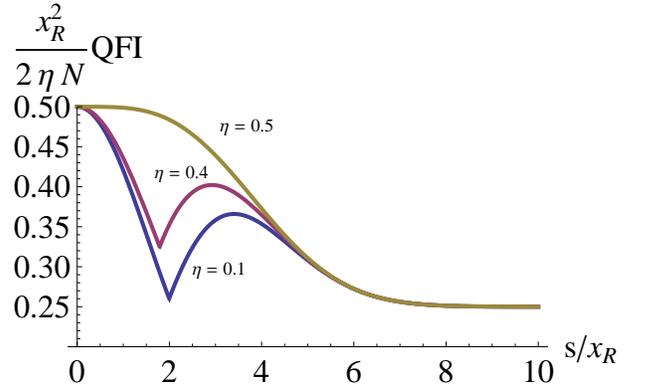}
\caption{ Ultimate precision bound for the estimation of the separation
between two point-like sources, measured in Rayleigh units. The plot shows
the ultimate quantum Fisher information per photon, for a Gaussian point
spread function. From bottom to top, we consider the following optical
attenuation $\protect\eta =0.1,0.4,$ and $0.5$. The corresponding functions $%
f_{\pm }$ are plotted in Fig.\ \ref{fig:fs} in the Appendix.}
\label{ultimateFIG}
\end{figure}

The upper bound in Eq.~(\ref{UB-new}) is proportional to the mean number of
collected photons $2\eta N$, according to the standard quantum limit. 
This property is directly inherited from the optimal estimation of a 
lossy bosonic channel.
As expected, the upper bound is inversely proportional to the square of the Rayleigh
length $\mathsf{x_{R}}$, in accordance to the fact that a smaller Rayleigh
length allows for higher resolution. Also note that the bound depends on the
two functions $f_{+}$ and $f_{-}$ which are the contributions of the
non-local source modes $c_{+}$ and $c_{-}$ to the quantum Fisher
information. In general, we expect that for $s\ll \mathsf{x_{R}}$ the
symmetric mode is almost insensitive to small variations of $s$, implying $%
f_{-}>f_{+}\simeq 0$. On the other hand, for $s\gg \mathsf{x_{R}}$ the two
sources decouple, yielding $f_{+}\simeq f_{-}$. Although these functions are
smooth, the maximum may occur in correspondence of a crossover (see Fig.\ \ref{ultimateFIG}).

%---

\textit{Achievability: optimal states.} Now we show that the upper bound
established in Theorem \ref{Theorem1} can in fact be achieved. Before
presenting optimal states saturating the bound, we derive the quantum Fisher
information for the case where the state of the light impinging on the image
screen takes the form
\begin{equation}
\rho _{a_{+}a_{-}}=\sum_{n,m}p_{nm}|n,m\rangle \langle n,m|\,,
\label{commute}
\end{equation}%
where $|n,m\rangle $ is a Fock state with $n$ photons in the symmetric mode
and $m$ photons in the antisymmetric one.

The quantum Fisher information for the parameter $s$ can be computed from $%
\mathrm{QFI}_{s}=\mathrm{Tr}\left( \mathcal{L}_{s}^{2}\rho \right) $, where $%
\mathcal{L}_{s}$ is the symmetric logarithmic derivative. For states as in
Eq.\ (\ref{commute}) and given that the modes $c_{\pm }$ emit $N_{\pm }$
mean photons each, we obtain 
\begin{equation}
\mathrm{QFI}_{s}=\langle \left( \partial _{s}\log {p}\right) ^{2}\rangle
+\eta N_{+}\epsilon _{+}^{2}+\eta N_{-}\epsilon _{-}^{2}~,  \label{optimal-d}
\end{equation}%
where $\langle \left( \partial _{s}\log {p}\right) ^{2}\rangle
=\sum_{nm}p_{nm}\left( \partial _{s}\log {p_{nm}}\right) ^{2}$. 
See Appendix \ref{App_C} for proof. Sources as in Eq.~(\ref{commute})
include thermal states and two-mode squeezed states, whose quantum Fisher
information is computed in Appendix \ref{sec:thermal_squeezed}.

The case of thermal states is particularly important since most natural
sources of light are thermal, especially in the setting of astronomical
observations. For two sources emitting $N$ mean thermal photons each we
obtain 
\begin{equation}
\mathrm{QFI}_{s}^{\mathrm{thermal}}=2\eta N\left[ \Delta k^{2}-\frac{\eta
N(1+\eta N)\gamma ^{2}}{(1+\eta N)^{2}-\delta ^{2}\eta ^{2}N^{2}}\right] \,.
\label{T-sources}
\end{equation}%
This result extends that of Ref.~\cite{Tsang}, which considered highly
attenuated incoherent sources, to the case of thermal sources of any
intensity. A comparison with Eq.\ (\ref{UB-new}) shows that thermal light is
always suboptimal for estimating the separation between the sources, apart
from the region $s\gg \mathsf{x_{R}}$, where we obtain $\mathrm{QFI}_{s}^{%
\mathrm{thermal}}\simeq \eta N\Delta k^{2}$. An interesting regime is that
of highly attenuated light ($\eta N\ll 1$) in which case we find $\mathrm{QFI%
}_{s}^{\mathrm{thermal}}\simeq \eta N\Delta k^{2}$ for all values of the
separation $s$.

From Eq.~(\ref{optimal-d}) it follows that the optimal states among
number-diagonal states are those maximizing $\langle \left( \partial
_{s}\log {p}\right) ^{2}\rangle $, which incidentally is the classical
Fisher information of the probability distribution $p_{nm}$~\cite{CCramer}.
This observation is explotied for proving the following result.%

\begin{theorem}
\label{Theorem2} For integer $2N$, the upper bound of Theorem \ref{Theorem1}
is saturated by sources $c_{+}$ and $c_{-}$ emitting the Fock state 
$|N_{+},N_{-}\rangle$ with $N_{+}+N_{-}=2N$. In particular, the optimal
state is either $|+\rangle :=|2N,0\rangle $ or $|-\rangle :=|0,2N\rangle $.
In terms of the original source modes, $c_{1}$ and $c_{2}$, these are the
entangled states 
\begin{equation}
|\pm \rangle =\frac{1}{2^{N}}\sum_{j=0}^{2N}\sqrt{{\binom{2N}{j}}}(\pm
1)^{2N-j}|j\rangle _{1}|2N-j\rangle _{2},
\end{equation}%
where $|k\rangle _{1,2}$ is a Fock state for $c_{1,2}$.
\end{theorem}

\textbf{Proof.}~~Since each mode $c_{\pm }$ is independently attenuated by a
attenuation factor $\eta _{\pm }$, the source state 
$|N_{+},N_{-}\rangle = (N_+! N_-!)^{-1/2}(c_{+}^\dag)^{N_+} (c_-^\dag)^{N_-} |0\rangle$
is mapped into an image state of the form (\ref{commute}) with $%
p_{nm}=p_{n}^{+}p_{m}^{-}$ and
\begin{equation}
p_{n}^{\pm }={\binom{N_{\pm }}{n}}\eta _{\pm }^{n}(1-\eta _{\pm })^{N_{\pm
}-n}\,.  \label{optimal_p}
\end{equation}%
For such a state we obtain 
\begin{align}
\langle \left( \partial _{s}\log {p}\right) ^{2}\rangle & =\langle \left(
\partial _{s}\log {p^{+}}\right) ^{2}\rangle +\langle \left( \partial
_{s}\log {p^{-}}\right) ^{2}\rangle  \\
& =\frac{\eta N_{+}\gamma ^{2}}{(1+\delta )(1-(1+\delta )\eta )}  \notag \\
& \hspace{0.5cm}+\frac{\eta N_{-}\gamma ^{2}}{(1-\delta )(1-(1-\delta )\eta )%
}\,.
\end{align}%
Inserting this result into Eq.\ (\ref{optimal-d}) we obtain that the Fock
state $|N_{+},N_{-}\rangle $ yields 
\begin{equation}
\mathrm{QFI}_{s}=\frac{\eta }{\mathsf{x_{R}}^{2}}\left(
N_{+}f_{+}+N_{-}f_{-}\right) ~.
\end{equation}%
The maximum of this quantity under the constraint $N_{+}+N_{-}=2N$ is
obtained by putting either $N_{+}=2N$ or $N_{+}=0$, hence saturating the
upper bound of Theorem \ref{Theorem2}. $\blacksquare $

%\vspace{0.5cm}

We remark that the optimal states in Theorem~\ref{Theorem2} have the same
form of the optimal states for the estimation of the loss parameter of a
bosonic channel \cite{BS2}. Following \cite{BS2}, optimal states for
non-integer $2N$ can be approximated by superposition of Fock states with
different photon numbers. Sources emitting photons in a two-mode squeezed
vacuum and sources of separable but quantum-correlated thermal light exhibits
features similar to the optimal states (see Appendix \ref{App_D2} and \ref{subsec:thermal_corr}).

\textit{Ultimate quantum Fisher information.} Having found a matching lower
bound implies that Eq.~(\ref{UB-new}) is in fact achievable and represents
the ultimate quantum Fisher information, optimized over the state of the
light emitted by the sources. It is clear that optimal states can be
explicitly engineered in all those scenarios where we can control the light
emitted by the sources, which is a typical case in microscopy.

For $s\gg \mathsf{x_{R}}$ the overlap $\delta $ between the image modes
becomes negligible, hence Eq.~(\ref{UB-new}) yields $\mathrm{QFI}_{s}\simeq
2\eta N\Delta k^{2}\sim 2\eta N\mathsf{x_{R}}^{-2}$. On the other hand, for
generic values of the separation $s$ and all values of the optical
attenuation $\eta $, we find $\mathrm{QFI}_{s}>2\eta N\Delta k^{2}$. This
means that the closer the sources are the better their distance can be
estimated. This counter-intuitive phenomenon is a \textit{super-resolution}
effect which appears at the sub-Rayleigh scale for entangled sources. We
have also found examples of quantum-correlated sources that are not
entangled (but discordant) which displays super-resolution at the
sub-Rayleigh scale (see Appendix \ref{subsec:thermal_corr}).

%---

The super-resolution is explicitly shown in the example of Fig.~\ref%
{ultimateFIG}, where we consider a Gaussian point spread function $\psi
(x)\sim \exp [{-x^{2}/(}4\mathsf{x_{R}}^{2})]$ with variance $\mathsf{x_{R}}%
^{2}$. In this case, $\Delta k^{2}=1/(4\mathsf{x_{R}}^{2})$ which yields $%
\lim_{s\gg \mathsf{x_{R}}}\mathrm{QFI}_{s}=2\eta N\Delta k^{2}=\eta N/(2%
\mathsf{x_{R}}^{2})$. Fig.~\ref{ultimateFIG} shows the ultimate (normalized)
quantum Fisher information per photon, i.e., $\mathsf{x_{R}}^{2}\mathrm{QFI}%
_{s}/(2\eta N)$, versus the dimensionless separation $s/\mathsf{x_{R}}$. The
upper bound has a maximum for finite $s$ in the sub-Rayleigh region. 
The maximum value of the quantum Fisher information per photon is $1/2$
which is reached for $s/\mathsf{x_{R}}\rightarrow 0$.

\textit{Optimal measurements at the sub-Rayleigh scale.} We now present a
sub-optimal measurement that is optimal for $s\lesssim \mathsf{x_{R}}$. We
consider sources emitting light in the optimal state as in Theorem~\ref%
{Theorem2}. We also consider a standard setting where 
the point spread function is symmetric around its center, i.e., $\psi
(x-y)=\psi (|x-y|)$. It follows that the image modes 
\begin{equation}
a_{\pm }^{\dag }=\frac{1}{\sqrt{2(1\pm \delta )}}\int \,dx\,\left[ \psi
(x+s/2)\pm \psi (x-s/2)\right] a_{x}^{\dag }\,,
\end{equation}%
are respectively even and odd functions of the coordinate $x$. %
We can hence consider a measurement able to distinguish the parity as, for
example, photo-detection of the Hermite-Gauss modes. (This kind of
measurement is optimal also in other settings~\cite{Tsang}. See also Ref.~%
\cite{Dispelling} for a more general approach).

Consider photon counting in the space of even and odd functions. The
probability of counting $n$ photons in even modes is given by $p_{n}^{+}$,
and the probability of $n$ photons in odd modes is given by $p_{n}^{-}$. The
(classical) Fisher information associated to this measurement is 
\begin{align}
F_{s}& =\left\langle \left( \partial _{s}\log {p^{+}}\right)
^{2}\right\rangle +\left\langle \left( \partial _{s}\log {p^{-}}\right)
^{2}\right\rangle  \\
& =\frac{\eta N_{+}\gamma ^{2}}{(1+\delta )(1-(1+\delta )\eta )}+\frac{\eta
N_{-}\gamma ^{2}}{(1-\delta )(1-(1-\delta )\eta )}\,.  \notag
\end{align}%
For $s\lesssim \mathsf{x_{R}}$, this non-adaptive measurement is optimal for
almost all values of $\eta $ (but for $\eta \sim 0.5$). For larger values of 
$s$ it fails to be optimal because it does not account for the fact that a
change in the value of the separation $s$ also implies a translation of the
image modes on the image screen.

\textit{Conclusions.} We have found the ultimate precision bound for
estimating the separation between two point-like sources using a 
linear-optical imaging system, considering arbitrary quantum states for the
sources. 
Although we have focused on the problem of estimating the separation between
two sources, our approach can be immediately extended to the problem of
estimating the location of a single source.

Our findings show that the separation between sources emitting
quantum-correlated light (entangled or discordant) can be super-resolved at
the sub-Rayleigh region. In particular, we have found the optimal entangled
states with this feature. 
Under optimal conditions one can increase the sub-Rayleigh quantum Fisher
information by a constant factor with respect of its value for separations
much larger than the Rayleigh length. 
In the sub-Rayleigh regime, we have shown that photon counting in the
symmetric and antisymmetric modes is an optimal measurement.

Another consequence of our findings is that the ultimate accuracy for any
linear-optical, far-field imaging system in the paraxial approximation
scales according to the standard quantum limit. While in principle it could still
be possible to beat the standard quantum limit, our results show that in
order to do so it is necessary to rely on a biased estimator for the source
separation, to consider non-point-like sources, or to employ a near-field,
non-linear, or non-paraxial imaging system.

%---

%---

\textit{Note added.} The specification of our general result of Eq.~(\ref%
{UB-new}) to the case of thermal sources [see Eq.~(\ref{T-sources})] has
been independently found by Nair and Tsang \cite{Nair}; these authors also
study tailored measurements that are almost optimal for estimating the
separation between two thermal sources.

\textit{Acknowledgments}.~We thank R. Nair, M. Tsang, K. Macieszczak, G.
Adesso and G. A. Durkin for their valuable comments. C.L. is also grateful
to R. Accardo and L. Iuorio, without whose support this work would not have
been possible.

\begin{widetext}

%%%%%%%%%%%%%%%%%%%%%%%%%%%%%%%%%%%%%%%%%%%%%%%%%%%%%%%%%%%%%%%%%%%%%%%%%%%%%%%%%%%%%%%%%%%%%%%%%%%%%%%%

\appendix

\section{A set of normal modes for the field on the image plane}\label{App_A}

In the main body of the paper we define the operators 
\begin{align}
a_1 & = \int \, dx \, \psi^*(x+s/2) \, a_x \, , \\
a_2 & = \int \, dx \, \psi^*(x-s/2) \, a_x \, , \\
\frac{ \partial a_1 }{\partial s} & = \int \, dx \, \frac{d \psi^*(x+s/2)}{d
s} \, a_x \, , \\
\frac{ \partial a_2 }{\partial s} & = \int \, dx \, \frac{d \psi^*(x-s/2)}{d
s} \, a_x \, ,
\end{align}
where $a_x$, $a_x^\dag$ are the canonical creation and annihilation
operators for the field at location $x$ on the image screen.

The operators $a_1$, $a_2$, $\frac{\partial a_1}{\partial s}$, $\frac{%
\partial a_1}{\partial s}$ do not define a set of canonical bosonic modes.
Following \cite{Tsang}, we define the operators $a_+$, $a_-$, $b_+$, $b_-$
as: 
\begin{align}
a_- & = \frac{a_1 - a_2}{\sqrt{2(1-\delta)}} \,, \\
a_+ & = \frac{a_1 + a_2}{\sqrt{2(1+\delta)}} \,, \\
b_{-} & = - \frac{\sqrt{2}}{\epsilon_-} \left[ \frac{\partial a_1}{\partial s%
} - \frac{\partial a_2}{\partial s} + \frac{\gamma}{\sqrt{2(1-\delta)}} \,
a_{-} \right] \, , \\
b_{+} & = - \frac{\sqrt{2}}{\epsilon_+} \left[ \frac{\partial a_1}{\partial s%
} + \frac{\partial a_2}{\partial s} - \frac{\gamma}{\sqrt{2(1+\delta)}} \,
a_{+} \right] \, ,
\end{align}
with 
\begin{align}
\delta & = \mathrm{Re} \int dx \psi^*(x+s/2) \psi(x-s/2) \, , \\
\epsilon_-^2 & = \Delta k^2 + \beta - \frac{\gamma^2}{1-\delta} \, , \\
\epsilon_+^2 & = \Delta k^2 - \beta - \frac{\gamma^2}{1+\delta} \, , \\
\beta & = \int dx \frac{d \psi(x+s/2)}{d x} \, \frac{d \psi(x-s/2)}{d x} \, ,
\\
\Delta k^2 & = \int dx \left| \frac{d \psi(x)}{d x} \right|^2 \, ,
\label{dk2_app} \\
\gamma & = \int dx \frac{d \psi(x)}{d x} \, \psi(x-s) = \frac{d \delta}{d s}
\, .
\end{align}
The modes $a_+$, $a_-$, $b_+$, $b_-$ satisfy canonical commutation relations
under the condition that the phase of the point-spread function $\psi$ is
constant, which is the case up to corrections of the second order in the
paraxial approximation \cite{Goodman}. In any case, as noted in \cite{Tsang}%
, a possible residual phase can be locally compensated.

%---

In terms of the canonical operators $a_\pm$, $b_\pm$ we then obtain 
\begin{equation}
\frac{ \partial a_1 }{ \partial s } \pm \frac{ \partial a_2 }{ \partial s }
= - \frac{\epsilon_\pm}{\sqrt{2}} \, b_\pm \pm \frac{\gamma}{\sqrt{%
2(1\pm\delta)}} \, a_\pm \, ,
\end{equation}
and 
\begin{equation}  \label{partial-a-supp}
\frac{\partial a_\pm}{\partial s} = - \frac{\epsilon_\pm}{2\sqrt{ 1 \pm
\delta }} \, b_\pm \, .
\end{equation}

%---

\section{Variance of the effective beam-splitter Hamiltonian}\label{app:variance}

In this Section we compute the variance %in Eq.\ (\ref{the_variance}).
\begin{align}
\langle \psi^{\prime}| \Delta^2 H^\mathrm{eff} | \psi^{\prime}\rangle & =
\langle \psi^{\prime}| (H^\mathrm{eff})^2 | \psi^{\prime}\rangle - \langle
\psi^{\prime}| H^\mathrm{eff} | \psi^{\prime}\rangle^2 \\
& = \langle \psi^{\prime}| \left( \omega_+ H_+^\mathrm{eff} + \omega_- H_-^%
\mathrm{eff} \right)^2 | \psi^{\prime}\rangle - \left( \langle
\psi^{\prime}| \omega_+ H_+^\mathrm{eff} + \omega_- H_-^\mathrm{eff} |
\psi^{\prime}\rangle \right)^2 \, ,
\end{align}
where 
\begin{equation}  \label{psi1-supp}
|\psi^{\prime}\rangle = \left( e^{-i \theta_+ H_+} e^{-i \theta_- H_-}
|\psi\rangle_{c_+c_-} |0\rangle_{v_+v_-} \right) |0\rangle_{b_+b_-} \, .
\end{equation}
is the state of the light at the image screen, together with the state of
the auxiliary modes $v_\pm$, $b_\pm$.

Here 
\begin{equation}
H_{\pm} = i \left( c_\pm^\dag v_\pm - v_\pm^\dag c_\pm \right) \, ,
\end{equation}
and 
\begin{equation}
H^\mathrm{eff} = \omega_+ H_+^\mathrm{eff} + \omega_- H_-^\mathrm{eff} \, ,
\end{equation}
with 
\begin{equation}
\omega_\pm H_\pm^\mathrm{eff} = i A_\pm \left( c_\pm^\dag v_\pm - c_\pm
v_\pm^\dag \right) + i B_\pm \left( c_\pm^\dag b_\pm - c_\pm^\dag b_\pm
\right) \, ,
\end{equation}
and $A_\pm = \frac{d\theta_\pm}{ds}$, $B_\pm = \frac{\epsilon_\pm}{2\sqrt{ 1
\pm \delta }}$.

We then obtain 
\begin{align}
\langle \psi^{\prime}| \Delta^2 H^\mathrm{eff} | \psi^{\prime}\rangle & =
\langle \psi^{\prime}| \left[ i A_+ \left( c_+^\dag v_+ - c_+ v_+^\dag
\right) + i A_- \left( c_-^\dag v_- - c_- v_-^\dag \right) \right.  \notag \\
& \hspace{1cm} \left. + i B_+ \left( c_+^\dag b_+ - c_+ b_+^\dag \right) + i
B_- \left( c_-^\dag b_- - c_- b_-^\dag \right) \right]^2 |
\psi^{\prime}\rangle  \notag \\
& \hspace{0.5cm} - \langle \psi^{\prime}| \left[ i A_+ \left( c_+^\dag v_+ -
c_+ v_+^\dag \right) + i A_- \left( c_-^\dag v_- - c_- v_-^\dag \right)
\right.  \notag \\
& \hspace{1cm} \left. + i B_+ \left( c_+^\dag b_+ - c_+ b_+^\dag \right) + i
B_- \left( c_-^\dag b_- - c_- b_-^\dag \right) \right] |
\psi^{\prime}\rangle^2 \, .
\end{align}

It follows from (\ref{psi1-supp}) that $b_\pm | \psi^{\prime}\rangle = 0$,
which implies 
\begin{align}
\langle \psi^{\prime}| \Delta^2 H^\mathrm{eff} | \psi^{\prime}\rangle & =
\langle \psi^{\prime}| \left[ i A_+ \left( c_+^\dag v_+ - c_+ v_+^\dag
\right) + i A_- \left( c_-^\dag v_- - c_- v_-^\dag \right) \right]^2
|\psi^{\prime}\rangle  \notag \\
& \hspace{0.5cm} + \langle \psi^{\prime}| \left[ i B_+ \left( c_+^\dag b_+ -
c_+ b_+^\dag \right) + i B_- \left( c_-^\dag b_- - c_- b_-^\dag \right) %
\right]^2 | \psi^{\prime}\rangle  \notag \\
& \hspace{0.5cm} - \langle \psi^{\prime}| \left[ i A_+ \left( c_+^\dag v_+ -
c_+ v_+^\dag \right) + i A_- \left( c_-^\dag v_- - c_- v_-^\dag \right) %
\right] | \psi^{\prime}\rangle^2 \\
& = \langle \psi^{\prime}| \left[ i A_+ \left( c_+^\dag v_+ - c_+ v_+^\dag
\right) + i A_- \left( c_-^\dag v_- - c_- v_-^\dag \right) \right]^2
|\psi^{\prime}\rangle  \notag \\
& \hspace{0.5cm} + \langle \psi^{\prime}| \left( B_+^2 c_+^\dag c_+ + B_-^2
c_-^\dag c_- \right) | \psi^{\prime}\rangle  \notag \\
& \hspace{0.5cm} - \langle \psi^{\prime}| \left[ i A_+ \left( c_+^\dag v_+ -
c_+ v_+^\dag \right) + i A_- \left( c_-^\dag v_- - c_- v_-^\dag \right) %
\right] | \psi^{\prime}\rangle^2 \, .
\end{align}

Since the operator $i A_+ \left( c_+^\dag v_+ - c_+ v_+^\dag \right) + i A_-
\left( c_-^\dag v_- - c_- v_-^\dag \right)$ commutes with the Hamiltonian $%
H_+ + H_-$, its expectation values on $|\psi^{\prime}\rangle = \left( e^{-i
\theta_+ H_+} e^{-i \theta_- H_-} |\psi\rangle_{c_+c_-} |0\rangle_{v_+v_-}
\right) |0\rangle_{b_+b_-}$ equal the expectations values on $%
|\psi,0\rangle:= |\psi\rangle_{c_+c_-} |0\rangle_{v_+v_-}$, which implies 
\begin{align}
\langle \psi^{\prime}| \Delta^2 H^\mathrm{eff} | \psi^{\prime}\rangle & =
\langle \psi,0| \left[ i A_+ \left( c_+^\dag v_+ - c_+ v_+^\dag \right) + i
A_- \left( c_-^\dag v_- - c_- v_-^\dag \right) \right]^2 |\psi,0\rangle 
\notag \\
& \hspace{0.5cm} + \langle \psi^{\prime}| \left( B_+^2 c_+^\dag c_+ + B_-^2
c_-^\dag c_- \right) | \psi^{\prime}\rangle  \notag \\
& \hspace{0.5cm} - \langle \psi,0| \left[ i A_+ \left( c_+^\dag v_+ - c_+
v_+^\dag \right) + i A_- \left( c_-^\dag v_- - c_- v_-^\dag \right) \right]
| \psi,0\rangle^2 \\
& = \langle \psi,0| \left( A_+^2 c_+^\dag c_+ + A_-^2 c_-^\dag c_- \right)
|\psi,0\rangle + \langle \psi^{\prime}| \left( B_+^2 c_+^\dag c_+ + B_-^2
c_-^\dag c_- \right) | \psi^{\prime}\rangle \, ,
\end{align}
that is, 
\begin{align}
\langle \psi^{\prime}| \Delta^2 H^\mathrm{eff} | \psi^{\prime}\rangle & =
\left( \frac{d \theta_+}{ds} \right)^2 \langle \psi | c_+^\dag c_+ | \psi
\rangle + \frac{\epsilon_+^2}{4(1+\delta)} \langle \psi^{\prime}| c_+^\dag
c_+ | \psi^{\prime}\rangle  \notag \\
& \hspace{0.5cm} + \left( \frac{d \theta_-}{ds} \right)^2 \langle \psi |
c_-^\dag c_- | \psi \rangle + \frac{\epsilon_-^2}{4(1-\delta)} \langle
\psi^{\prime}| c_-^\dag c_- |\psi^{\prime}\rangle \, .  \label{the_variance}
\end{align}

It is important to remark that in this last expression for the variance $%
\langle \psi^{\prime}| \Delta^2 H^\mathrm{eff} | \psi^{\prime}\rangle$,
there are no cross-terms coupling the operators $c_+$, $c_+^\dag$ with the
operators $c_-$, $c_-^\dag$. This implies that the upper bound on the
quantum Fisher information is the sum of two independent terms.

Assume that the source $c_\pm$ emits $N_\pm$ mean photons, that is, $\langle
\psi | c_\pm^\dag c_\pm | \psi \rangle = N_\pm$, which in turn implies $%
\langle \psi^{\prime}| c_\pm^\dag c_\pm | \psi^{\prime}\rangle = \eta_\pm
N_\pm = (1\pm\delta)\eta N_\pm$. Then we obtain the following upper bound on
the quantum Fisher information: 
\begin{align}
\mathrm{QFI}_s & \leq 4 N_+ \left[ \left( \frac{d \theta_+}{ds} \right)^2 + 
\frac{\eta \epsilon_+^2}{4}\right] + 4 N_- \left[ \left( \frac{d \theta_-}{ds%
} \right)^2 + \frac{\eta \epsilon_-^2}{4} \right] \\
& = N_+ \left[ \frac{ \eta \gamma^2}{(1+\delta)(1-(1+\delta)\eta)} + \eta
\epsilon_+^2 \right] + N_- \left[ \frac{\eta \gamma^2}{(1-\delta)(1-(1-%
\delta)\eta)} + \eta \epsilon_-^2 \right] \, .  \label{bound-diff}
\end{align}
Given a constraint of $N_+ + N_- = 2N$ total mean photons emitted by the
sources, the maximum of the right hand side is obtained putting either $N_+
= 2N$ and $N_- = 0$ or $N_+ = 0$ and $N_- = 2N$. We then obtain 
\begin{equation}  \label{UB}
\mathrm{QFI}_s \leq 2 \eta N \max\left\{ \frac{ \gamma^2}{%
(1+\delta)(1-(1+\delta)\eta)} + \epsilon_+^2 \, , \, \frac{\gamma^2}{%
(1-\delta)(1-(1-\delta)\eta)} + \epsilon_-^2 \right\} \, .
\end{equation}

As an example, consider the case of a Gaussian point-spread function, $%
\psi(x) \sim \exp{\ \left( - \frac{x^2}{4 \mathsf{x_R}^2} \right) }$. The
quantities 
\begin{equation}  \label{fp}
f_+ := \mathsf{x_R}^2 \left[ \frac{ \gamma^2}{(1+\delta)(1-(1+\delta)\eta)}
+ \epsilon_+^2 \right]
\end{equation}
and 
\begin{equation}  \label{fm}
f_- := \mathsf{x_R}^2 \left[ \frac{\gamma^2}{(1-\delta)(1-(1-\delta)\eta)} +
\epsilon_-^2 \right]
\end{equation}
are plotted versus the $s/\mathsf{x_R}$ in Fig.\ \ref{fig:fs} for different
values of $\eta$. As the figures show, which is the maximum between $f_+$
and $f_-$ depend both on $\eta$ and $s$.

\begin{figure}[tbp]
%\centering
\begin{minipage}{3.5in}
\includegraphics[width=1\textwidth]{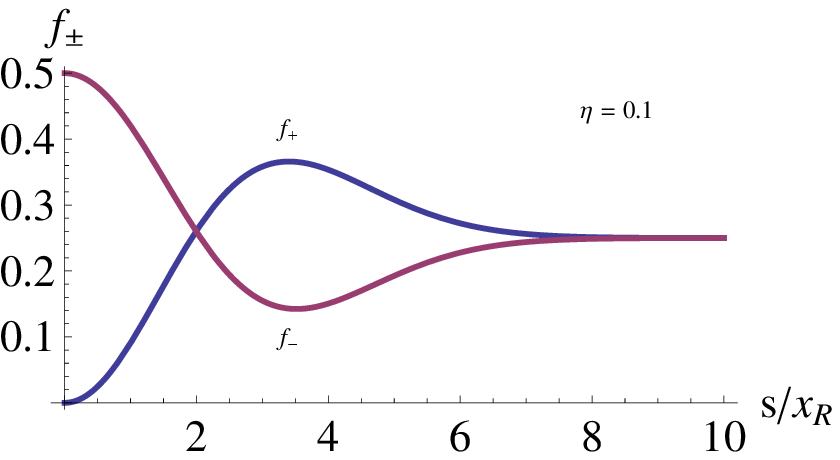}
\end{minipage}\\%\newline
\begin{minipage}{3.5in}
\includegraphics[width=1\textwidth]{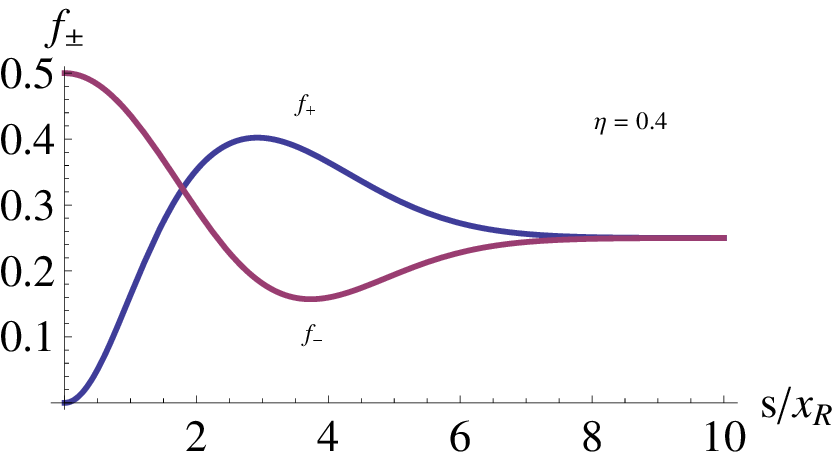}
\end{minipage}\\%\newline
\begin{minipage}{3.5in}
\includegraphics[width=1\textwidth]{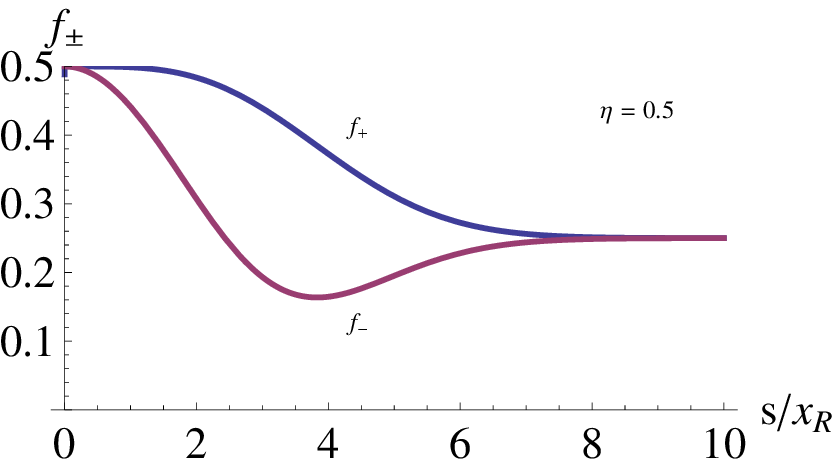}
\end{minipage}
\caption{The quantities $f_\pm$ defined in Eqs.\ (\protect\ref{fp})-(\protect
\ref{fm}) plotted versus the separation $s/\mathsf{x_R}$ for different
values of $\protect\eta$. From top to bottom, $\protect\eta = 0.1, 0.4, 0.5$.
}
\label{fig:fs}
\end{figure}

%---

\section{Quantum Fisher information for number-diagonal states}\label{App_C}

In this section we compute the quantum Fisher information for
number-diagonal states of the form 
\begin{equation}  \label{commute-supp}
\rho_{a_+a_-} = \sum_{n,m} p_{nm} | n ; m \rangle \langle n ; m |\, ,
\end{equation}
where $| n ; m \rangle = \left( n! m! \right)^{-1/2} {a_+^\dag}^n {a_-^\dag}%
^m |0\rangle$ denotes a Fock state with $n$ photons in the symmetric mode
and $m$ photons in the anti-symmetric one.

The quantum Fisher information for the parameter $s$ can be computed by
applying the formula $\mathrm{QFI}_s = \mathrm{Tr} \left( \mathcal{L}_s^2
\rho \right)$, where 
\begin{equation}
\mathcal{L}_s =
\sum_{n,n^{\prime},m,m^{\prime}|p_{nm}+p_{n^{\prime}m^{\prime}} > 0} \left( 
\frac{2}{p_{nm}+p_{n^{\prime}m^{\prime}}} \right) |n;m\rangle \langle n;m| 
\frac{\partial \rho}{\partial s} | n^{\prime};m^{\prime}\rangle \langle
n^{\prime};m^{\prime}|
\end{equation}
is the symmetric logarithmic derivative.

The derivative of $\rho$ with respect to $s$ reads 
\begin{align}
\partial_s \rho & = \sum_{n,m} \left( \partial_s p_{nm} \right) \,
|n;m\rangle \langle n;m| + p_{nm} \partial_s \left( |n;m\rangle \langle n;m|
\right)  \label{Leibnitz} \\
& = \sum_{n,m} \left( \partial_s p_{nm} \right) \, |n;m\rangle \langle n;m|
+ p_{nm} \partial_s \left( \frac{{a_+^\dag}^n}{\sqrt{n!}} \frac{{a_-^\dag}^m%
}{\sqrt{m!}} |0\rangle \langle 0| \frac{a_+^n}{\sqrt{n!}} \frac{a_-^m}{\sqrt{%
m!}} \right) \\
& = \sum_{n,m} \left( \partial_s p_{nm} \right) \, |n;m\rangle \langle n;m| 
\notag \\
& \hspace{0.5cm} + p_{nm} \frac{n}{\sqrt{n!}} \, \frac{\partial {a_+^\dag} }{%
\partial s} \, {a_+^\dag}^{n-1} |0;m\rangle \langle n;m| + p_{nm} \frac{n}{%
\sqrt{n!}} \, |n;m\rangle \langle 0;m| {a_+}^{n-1} \frac{\partial {a_+} }{%
\partial s}  \notag \\
& \hspace{0.5cm} + p_{nm} \frac{m}{\sqrt{m!}} \, \frac{\partial {a_-^\dag} }{%
\partial s} \, {a_-^\dag}^{m-1} |n;0\rangle \langle n;m| + p_{nm} \frac{m}{%
\sqrt{m!}} \, |n;m\rangle \langle n;0| {a_-}^{m-1} \frac{\partial {a_-} }{%
\partial s} \\
& = \sum_{n,m} \left( \partial_s p_{nm} \right) \, |n;m\rangle \langle n;m| 
\notag \\
& \hspace{0.5cm} + p_{nm} \sqrt{n} \left( \frac{\partial {a_+^\dag} }{%
\partial s} |n-1;m\rangle \langle n;m| + |n;m\rangle \langle n-1;m| \frac{%
\partial {a_+} }{\partial s} \right)  \notag \\
& \hspace{0.5cm} + p_{nm} \sqrt{m} \left( \frac{\partial {a_-^\dag} }{%
\partial s} |n;m-1\rangle \langle n;m| + |n;m\rangle \langle n;m-1| \frac{%
\partial {a_-} }{\partial s} \right) \, .
\end{align}

Now we apply Eq.\ (\ref{partial-a-supp}). Putting $|n-1,1;m\rangle =
b_+^\dag |n-1;m\rangle$ and $|n;m-1,1\rangle = b_-^\dag |n;m-1\rangle$ we
obtain 
\begin{align}
\partial_s \rho & = \sum_{n,m} \left( \partial_s p_{nm} \right) \,
|n;m\rangle \langle n;m|  \notag \\
& \hspace{0.5cm} - \frac{p_{nm} \sqrt{n} \, \epsilon_+}{2\sqrt{1+\delta}}
\left( |n-1,1;m\rangle \langle n;m| + |n;m\rangle \langle n-1,1;m| \right) 
\notag \\
& \hspace{0.5cm} - \frac{ p_{nm} \sqrt{m} \, \epsilon_-}{2\sqrt{1-\delta}}
\left( |n;m-1,1\rangle \langle n;m| + |n;m\rangle \langle n;m-1,1| \right)
\, ,  \label{drho}
\end{align}
from which we compute the symmetric logarithmic derivative 
\begin{align}
\mathcal{L}_s & = \sum_{n,m} \left( \partial_s \log{p_{nm}} \right) \,
|n;m\rangle \langle n;m|  \notag \\
& \hspace{0.5cm} - \frac{\sqrt{n} \, \epsilon_+}{\sqrt{1+\delta}} \left(
|n-1,1;m\rangle \langle n;m| + |n;m\rangle \langle n-1,1;m| \right)  \notag
\\
& \hspace{0.5cm} - \frac{ \sqrt{m} \, \epsilon_-}{\sqrt{1-\delta}} \left(
|n;m-1,1\rangle \langle n;m| + |n;m\rangle \langle n;m-1,1| \right) \, .
\end{align}

Assuming that each source mode $c_\pm$ emits $N_\pm$ mean photons, then the
quantum Fisher information reads 
\begin{equation}  \label{optimal-d_app}
\mathrm{QFI}_s = \langle \left( \partial_s \log{p} \right)^2 \rangle + \eta
N_+ \epsilon_+^2 + \eta N_- \epsilon_-^2 \, ,
\end{equation}
with $\langle \left( \partial_s \log{p} \right)^2 \rangle = \sum_{nm} p_{nm}
\left( \partial_s \log{p_{nm}} \right)^2$.

%---

\section{Special cases: thermal, correlated, and squeezed sources}\label{sec:thermal_squeezed}

\subsection{Thermal sources}\label{App_D1}

Let us first assume that the modes $c_1$, $c_2$ emit thermal monochromatic
light at a given temperature, with $N$ mean photons each, then the modes $%
c_\pm$ will be thermal too and at the same temperature. 
It follows that the modes $a_\pm$ on the image screen are also thermal, but with mean
photons $M_\pm = \eta (1\pm \delta) N$, respectively.

The state of the two image modes has the form 
\begin{align}
\rho_{a_+a_-} = \frac{1}{M_+ + 1} \frac{1}{M_- + 1} \sum_{n,m} \left( \frac{%
M_+}{ M_+ + 1} \right)^n \left( \frac{M_-}{ M_- + 1} \right)^m |n;m\rangle
\langle n;m| \, .
\end{align}
Since this is a number-diagonal state as in Eq.\ (\ref{commute-supp}), we
can apply Eq.\ (\ref{optimal-d_app}). A straightforward calculation then yields 
\begin{equation}
\langle \left( \partial_s \log{p} \right)^2 \rangle = 2 \eta N \left[ \frac{%
\gamma^2}{2(1+\delta)(1+(1+\delta)\eta N)} + \frac{\gamma^2}{%
2(1-\delta)(1+(1-\delta)\eta N)} \right] \, ,
\end{equation}
and 
\begin{align}
\mathrm{QFI}_s & = \langle \left( \partial_s \log{p} \right)^2 \rangle +
2\eta N \left[ \Delta k^2 - \frac{\gamma^2}{2(1+\delta)} - \frac{\gamma^2}{%
2(1-\delta)} \right]  \label{QFthermal} \\
& = 2 \eta N \left[ \frac{\gamma^2}{2(1+\delta)(1+(1+\delta)\eta N)} + \frac{%
\gamma^2}{2(1-\delta)(1+(1-\delta)\eta N)} \right] + 2\eta N \left[ \Delta
k^2 - \frac{\gamma^2}{2(1+\delta)} - \frac{\gamma^2}{2(1-\delta)} \right] \\
& = 2\eta N \left[ \Delta k^2 - \frac{\eta N ( 1 + \eta N ) \gamma^2}{( 1 +
\eta N )^2 - \delta^2 \eta^2 N^2 } \right] \, .
\end{align}

In the semiclassical limit, $\eta N \gg 1$, the quantum Fisher information
per photon reads $\lim_{\eta N \gg 1} \mathrm{QFI}_s/(2\eta N) = \Delta k^2
- \frac{\gamma^2}{2(1+\delta)} - \frac{\gamma^2}{2(1-\delta)}$. As an
example, consider the Gaussian point-spread function, $\psi(x) \sim \exp{\
\left( - \frac{x^2}{4 \mathsf{x_R}^2} \right) }$. This is shown in Fig.\ \ref%
{fig:thermal}, where one can see that for bright thermal sources the quantum
Fisher information vanishes for $s \ll \mathsf{x_R}$ --- a typical classical
feature dubbed the ``Rayleigh's curse'' in \cite{Tsang}. We note that the
same result for thermal sources has been independently obtained by \cite%
{Nair}.

\begin{figure}[t]
\centering
\includegraphics[width=0.5\textwidth]{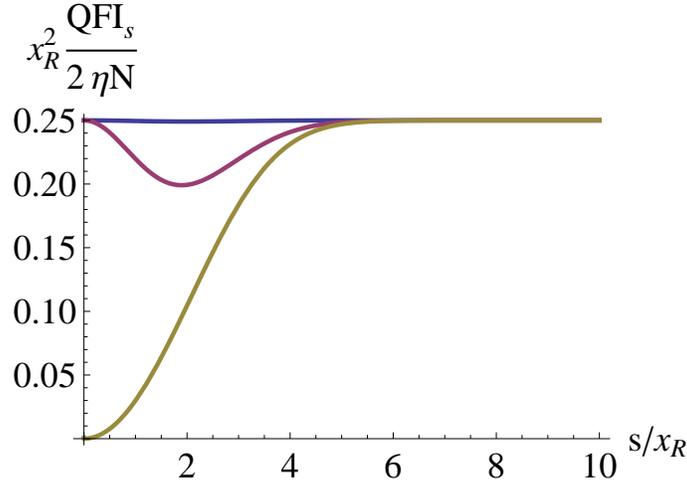}
\caption{ Quantum Fisher information for the estimation of the separation
between two point-like sources emitting thermal light. The plot shows the
quantum Fisher information per photon for a Gaussian point-spread function
with variance $\mathsf{x_R}^2$. From top to bottom $\protect\eta N = 0.01, 1$
and the semiclassical limit $\protect\eta N \to \infty$. }
\label{fig:thermal}
\end{figure}

%---

\subsection{Two-mode squeezed sources}\label{App_D2}

We now consider the case of two sources emitting a continuous-variable
quadrature-entangled state (two-mode squeezed vacuum) of the form \cite%
{WeeRMP,FOP} 
\begin{equation}
|\xi\rangle = \exp{\left[ \xi \left( c_1^\dag c_2^\dag - c_1 c_2 \right) %
\right]} |0\rangle \, .
\end{equation}
In terms of the modes $c_\pm$, this reads 
\begin{equation}
|\xi\rangle = \exp{\left[ \frac{\xi}{2} \left( {c_+^\dag}^2 - c_+^2 \right) %
\right]} \otimes \exp{\left[ -\frac{\xi}{2} \left( {c_-^\dag}^2 - c_-^2
\right) \right]} |0\rangle \, ,
\end{equation}
which describes the direct product of two independent squeezed vacua \cite%
{WeeRMP,FOP}. Each of these modes is independently attenuated when it enters
the optical imaging system, that is, the state of the modes $a_\pm$ on the
image screen is that of two independent attenuated squeezed vacua.

To describe these states it is convenient to use the covariance matrix of
the quadrature operators, $X_\pm = (c_\pm + c_\pm^\dag)/\sqrt{2}$ and $P_\pm
= -i(c_\pm - c_\pm^\dag)/\sqrt{2}$. In our case the first moments of the
quadrature operators vanish, and the covariance matrix reads 
\begin{align}
V_{c_+,c_-} & = \langle \xi| \left( 
\begin{array}{cc}
X_+^2 & \frac{X_+ P_+ + P_+ X_+}{2} \\ 
\frac{X_+ P_+ + P_+ X_+}{2} & P_+^2%
\end{array}
\right) \oplus \left( 
\begin{array}{cc}
X_-^2 & \frac{X_- P_- + P_- X_-}{2} \\ 
\frac{X_- P_- + P_- X_-}{2} & P_-^2%
\end{array}
\right) |\xi\rangle \\
& = \left( 
\begin{array}{cc}
\frac{1}{2} \, e^{2\xi} & 0 \\ 
0 & \frac{1}{2} \, e^{-2\xi}%
\end{array}
\right) \oplus \left( 
\begin{array}{cc}
\frac{1}{2} \, e^{-2\xi} & 0 \\ 
0 & \frac{1}{2} \, e^{2\xi}%
\end{array}
\right) \, .
\end{align}

The covariance matrix describing the image modes is obtained by attenuating
the modes by factors $\eta_\pm = (1 \pm \delta)\eta$: 
\begin{align}
V_{a_+,a_-} = \left( 
\begin{array}{cc}
\frac{\eta_+}{2} \, e^{2\xi} + \frac{1-\eta_+}{2} & 0 \\ 
0 & \frac{\eta_+}{2} \, e^{-2\xi} + \frac{1-\eta_+}{2}%
\end{array}
\right) \oplus \left( 
\begin{array}{cc}
\frac{\eta_-}{2} \, e^{-2\xi} + \frac{1-\eta_-}{2} & 0 \\ 
0 & \frac{\eta_-}{2} \, e^{2\xi} + \frac{1-\eta_-}{2}%
\end{array}
\right) \, .
\end{align}
We can re-parameterize this covariance matrix in terms of the variables $%
T_\pm$, $r_\pm$ as 
\begin{align}
V_{a_+,a_-} = \left( 
\begin{array}{cc}
e^{2r_+}\left( T_+ + \frac{1}{2} \right) & 0 \\ 
0 & e^{-2r_+}\left( T_+ + \frac{1}{2} \right)%
\end{array}
\right) \oplus \left( 
\begin{array}{cc}
e^{-2r_-}\left( T_- + \frac{1}{2} \right) & 0 \\ 
0 & e^{2r_-}\left( T_- + \frac{1}{2} \right)%
\end{array}
\right) \, ,
\end{align}
with 
\begin{equation}
T_\pm = \frac{1}{2} \left( \sqrt{ \eta_\pm^2 + (1-\eta_\pm)^2 + 2 \eta_\pm
(1-\eta_\pm) \cosh{(2\xi)}} - 1 \right)
\end{equation}
and 
\begin{equation}
r_\pm = \frac{1}{2}\sinh^{-1}{\left( \frac{\eta_\pm \sinh{(2\xi)}}{2 T_\pm +
1} \right)} \, .
\end{equation}

%---

Using this parameterization, the state of each image mode is described as
being the result of applying a squeezing transformation to a thermal state.
That is, the state of the modes $a_+$, $a_-$ has the form. 
\begin{equation}
\rho_{a_+a_-} = \rho_+ \otimes \rho_- \, ,
\end{equation}
where 
\begin{equation}
\rho_\pm = \frac{1}{T_\pm + 1} \sum_{n=0}^\infty \left( \frac{T_\pm}{T_\pm +
1} \right)^n \, |e_n \rangle_\pm \langle e_n| = \sum_{n=0}^\infty p_n^\pm \,
|e_n \rangle_\pm \langle e_n|
\end{equation}
and 
\begin{equation}
|e_n \rangle_\pm = e^{-i K_\pm} \, |n\rangle_\pm = \exp{\left[ \pm \frac{%
r_\pm}{2} \left( {a_\pm^\dag}^2 - a_\pm^2 \right) \right]} \, |n\rangle_\pm 
\end{equation}
is a squeezed $n$-photon Fock state.

Therefore, the state $\rho_{a_+a_-}$ is not diagonal in the number basis,
but in the basis $\{ |e_n \rangle_\pm \}$ that is unitarily related to the
latter by the action of the unitary $e^{-i K_\pm}$. The derivative of $%
\rho_\pm$ with respect to $s$ reads 
\begin{align}
\partial_s \rho_\pm & = \sum_n \left( \partial_s p_n^\pm \right) |e_n
\rangle_\pm \langle e_n| + p_n^\pm \partial_s \left( |e_n \rangle_\pm
\langle e_n| \right) \\
& = \sum_n \left( \partial_s p_n^\pm \right) |e_n \rangle_\pm \langle e_n| +
p_n ^\pm \, e^{-i K_\pm} \partial_s \left( |n \rangle_\pm \langle n| \right)
e^{i K_\pm}  \notag \\
& \hspace{0.5cm} + p_n^\pm \left[ \partial_s \left( e^{-iK_\pm} \right) |n
\rangle_\pm \langle e_n| + |e_n \rangle_\pm \langle n| \partial_s \left(
e^{iK_\pm} \right) \right] \, .
\end{align}

Comparing with (\ref{Leibnitz}), now we have one extra term accounting for
the derivative of $e^{-i K_\pm}$ with respect to $s$. From the
Baker--Campbell--Hausdorff formula we obtain 
\begin{equation}
e^{i K_\pm} \partial_s e^{-i K_\pm} = - i A_\pm K_\pm + \frac{B_\pm}{2}
\left( a_\pm^\dag b_\pm^\dag - a_\pm b_\pm \right) \sinh{r_\pm} + \frac{B_\pm%
}{2} \left( a_\pm^\dag b_\pm - a_\pm b_\pm^\dag \right) \left( \cosh{r_\pm}
- 1 \right) \, ,
\end{equation}
with 
\begin{equation}
A_\pm = \frac{ \partial \log{r_\pm } }{\partial s}
\end{equation}
and 
\begin{equation}
B_\pm = \frac{\epsilon_\pm}{\sqrt{ 1 \pm \delta }} \, .
\end{equation}
This allows us to compute the quantum Fisher information. We obtain: 
\begin{multline}
\mathrm{QFI}_s = \left\langle \left( \frac{\partial \log{p^+}}{\partial s}
\right)^2 \right\rangle + \left\langle \left( \frac{\partial \log{p^-}}{%
\partial s} \right)^2 \right\rangle  \notag \\
+ \eta \left( \cosh{2\xi}-1 \right) \left( \Delta k^2 - \frac{\gamma^2}{%
2(1+\delta)} - \frac{\gamma^2}{2(1-\delta)} \right)  \notag \\
+ \frac{2(2T_+ + 1)^2}{2T_+^2 + 2T_+ +1} \, \frac{\partial r_+}{\partial s}
+ \frac{2(2T_- + 1)^2}{2T_-^2 + 2T_- +1} \, \frac{\partial r_-}{\partial s}
\, .
\end{multline}

%----

For $s \gg 1$ we have $\mathrm{QFI}_s \simeq \eta (\cosh{2\xi} - 1 ) \Delta
k^2$, where $\eta (\cosh{2\xi} - 1 )$ is the mean photon number impinging on
the image screen.

As an example, Fig.\ \ref{fig:tmsv} shows the quantum Fisher information per
photon impinging on the image screen versus $s/\mathsf{x_R}$, for the
Gaussian point-spread function. For small values of $\eta$ the light is
highly attenuated and we observed the same behavior as highly attenuated
incoherent sources, while for small $\eta$ and relatively large squeezing $%
\xi$ the state becomes effectively semiclassical and manifests the classical
``Rayleigh's curse'' \cite{Tsang}. On the other hand, for larger value of $%
\eta$ we observe a phenomenon of super-resolution for sub-Rayleigh
distances, similarly to the optimal entangled sources obtained in the main
body of this paper.

\begin{figure}[t]
\centering
\includegraphics[width=0.5\textwidth]{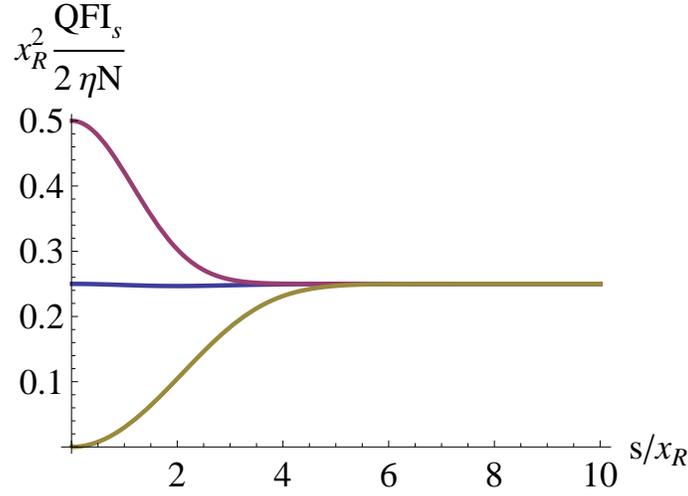}
\caption{ Quantum Fisher information for the estimation of the separation $s$
between two entangled sources emitting a two-mode squeezed vacuum state. The
plot show the quantum Fisher information per photon impinging on the image
screen for the case of a Gaussian point-spread function with variance $%
\mathsf{x_R}^2$. From top to bottom: $\protect\xi = 0.1$, $\protect\eta = 0.5
$; $\protect\xi = 1$, $\protect\eta = 0.01$; $\protect\xi = 10$, $\protect%
\eta = 0.1$. }
\label{fig:tmsv}
\end{figure}

%---

\subsection{Correlated thermal sources}\label{subsec:thermal_corr}

Let us considered two sources $c_1$, $c_2$ emitting light in a Gaussian
state with zero mean and covariance matrix 
\begin{align}
V_{c_1,c_2} = \left( 
\begin{array}{cccc}
N + \frac{1}{2} & 0 & w N & 0 \\ 
0 & N + \frac{1}{2} & 0 & w N \\ 
w N & 0 & N + \frac{1}{2} & 0 \\ 
0 & w N & 0 & N + \frac{1}{2}%
\end{array}
\right) \, .
\end{align}
For $w \leq 1$, this is a correlated thermal state. Such a state is always
separable but has non-zero discord. Its quantum discord can be computed
exactly according to the results of Ref.~\cite{D0}, and it coincides with
its Gaussian discord~\cite{D1,D2}.

Expressing the state in terms of the non-local source modes $c_+$ and $c_-$,
the covariance matrix reads 
\begin{align}
V_{c_+,c_-} = \left( 
\begin{array}{cccc}
N_+ + \frac{1}{2} & 0 & 0 & 0 \\ 
0 & N_+ + \frac{1}{2} & 0 & 0 \\ 
0 & 0 & N_- + \frac{1}{2} & 0 \\ 
0 & 0 & 0 & N_- + \frac{1}{2}%
\end{array}
\right) \, ,
\end{align}
which represents the product of two thermal states with different mean
photon number $N_+ = (1+w) N $ and $N_- = (1-w) N$.

To compute the quantum Fisher information associated to these sources, we
can proceed as we have done in Section \ref{sec:thermal_squeezed} for the
case of thermal sources emitting the same mean photons. We then obtain 
\begin{equation}
\mathrm{QFI}_s = \eta N_+ \left[ \delta k^2 - \beta - \frac{ \eta N_+
\gamma^2}{1+(1+\delta)\eta N_+} \right] + \eta N_- \left[ \delta k^2 + \beta
- \frac{ \eta N_- \gamma^2}{1+(1-\delta)\eta N_-} \right] \, .
\end{equation}

As an example, let us consider the regime of highly attenuated light, $\eta
N_\pm \ll 1$. In this limit the quantum Fisher information reads 
\begin{equation}
\mathrm{QFI}_s \simeq \eta N_+ \left( \delta k^2 - \beta \right) + \eta N_-
\left( \delta k^2 + \beta \right) = 2 \eta N \left( \delta k^2 - w\beta
\right)
\end{equation}
and yields a phenomenon of super-resolution for same value of $w$. For
example, Fig.\ \ref{fig:thcorr} shows the quantum Fisher information versus
the separation for the case of a Gaussian point spread function, yielding
super-resolution for $w < 0$.

\begin{figure}[t]
\centering
\includegraphics[width=0.5\textwidth]{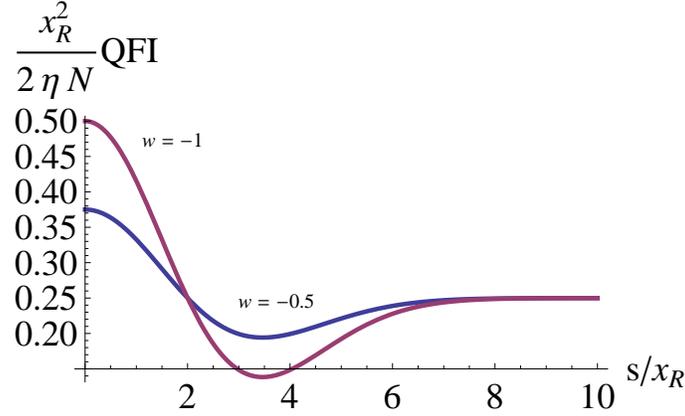}
\caption{ Quantum Fisher information for the estimation versus the
separation for sources emitting a two-mode correlated thermal state, for $%
w=-0.5$ and $w=-1$. These correlated sources allow for super-resolution at
the sub-Rayleigh scale. }
\label{fig:thcorr}
\end{figure}

%%%%%%%%%%%%%%%%%%%%%%%%%%%%%%%%%%%%%%%%%%%%%%%%%%%%%%%%%%%%%%%%%%%%%%%%%%%%%%%%%%%%%%%%%%%%%%%%%%%%%%%%

\end{widetext}


\begin{thebibliography}{99}
\bibitem{NOTAnew}
To derive this transformation rule, 
consider a single photon emitted in one of the source mode, say $c_1$.
The optical system transmits this photon with probability $\eta$, and
transfers it to an environmental mode $v_1$ with probability $1-\eta$, i.e.,
$$
|\psi\rangle = c_1^\dag |0\rangle \to  
\left( \sqrt{\eta} \, a_1^\dag + \sqrt{1-\eta} \, v_1^\dag \right) |0\rangle \, .
$$
\bibitem{NOTA1} The imaging system is characterized by the Fresnel number $%
\mathcal{F} = \ell / \mathsf{x_R}$, where $\ell$ is the size of the source
and $\mathsf{x_R}$ is the Rayleigh length. For $\mathcal{F} \ll 1$, the
imaging system operates in the far-field regime, in which light is
attenuated by a factor $\eta \simeq \mathcal{F}$ \cite{Shapiro,mio}. By
definition, a point-like source is such that its size is much smaller than
the typical size of the imaging system, which in our case is the Rayleigh
length. It follows that the optical system operates in the far-field regime
and the light emitted by the point source is attenuated by a factor $\eta
\simeq \mathcal{F} \ll 1$.
\end{thebibliography}
\end{document}